\documentclass[12pt]{iopart}
\usepackage{dcolumn}
\usepackage{bm}
\usepackage{graphicx}
\usepackage{subfigure}
 
\begin{document}
\title{Cooperativity-Driven Singularities in Asymmetric  Exclusion}
\author{Alan Gabel$^1$ and S. Redner$^1$}
\address{$^1$Center for Polymer Studies and Department of
  	Physics, Boston University, Boston, Massachusetts 02215, USA}
\ead{agabel@bu.edu, redner@bu.edu}

\begin{abstract}
  We investigate the effect of cooperative interactions on the asymmetric
  exclusion process, which causes the particle velocity to be an increasing
  function of the density.  Within a hydrodynamic theory, initial density
  upsteps and downsteps can evolve into: (a) shock waves, (b) continuous
  compression or rarefaction waves, or (c) a mixture of shocks and continuous
  waves.  These unusual phenomena arise because of an inflection point in the
  current versus density relation.  This anomaly leads to a group velocity
  that can either be an increasing or a decreasing function of the density on
  either side of the inflection point, a property that underlies these
  localized wave singularities.
\end{abstract}

\maketitle

\section{Introduction}

The asymmetric exclusion process (ASEP)~\cite{SZ95,D98,S00,BE07,KRB10}
represents an idealized description of transport in crowded one-dimensional
systems, such as traffic~\cite{PS99,AS00,SCN10}, ionic
conductors~\cite{R77}, and RNA transcription~\cite{MGB68}.  In the
ASEP, each site is either vacant or occupied by a single particle that can
hop at a fixed rate to a vacant right neighbor~\cite{SZ95,D98,S00,BE07}.
Although simply defined, this model has rich transport properties: for
example, density heterogeneities can evolve into rarefaction or shock
waves~\cite{BE07}, while an open system, with input at one end and output at
the other, exhibits a variety of phases as a function of the input/output
rates~\cite{K91,DDM92,SD93}.

\begin{figure}[htb]
\begin{center}
\includegraphics[width=0.35\textwidth]{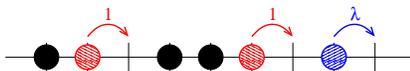}
\caption{Cooperative exclusion.  A ``pushed'' particle (red) --- one whose
  left neighbor is occupied --- can hop to a vacant right neighbor with rate
  1, while an isolated particle (blue) hops to a vacancy with rate
  $\lambda$.}
  \label{model}
\end{center}
\end{figure}

A fundamental property of the ASEP is the relation $J(\rho)=\rho(1-\rho)$
between the current $J$ and density $\rho$.  Because each site is occupied by
at most one particle, the average particle velocity $v=J/\rho$ is a
decreasing function of the density.  In this work, we investigate a {\em
  cooperative exclusion} (CE) model in which the velocity can \emph{increase}
with density.  This cooperativity leads to unexpected features in the
evolution of initial density heterogeneities.  Such cooperativity occurs, for
example, when ants emit pheromones that help guide fellow ants along a
trail~\cite{BAA02}.  Another example are multiple buses that follow a fixed
route.  The leading bus picks up more passengers so that the next bus moves
faster, which causes clustering of buses during peak travel
times~\cite{LEC98}.  At the microscopic level, molecular motors can work
together to pull a load that is too large for a single motor~\cite{C06}.
Cooperativity has even been proposed as a basis for organic
superconductors~\cite{L64}.

The notion of cooperative interactions that counterbalance the fundamental
excluded-volume interaction is implicit in Ref.~\cite{AS00}, as well as
in~\cite{FGRS02, BGRS02}.  These latter publications investigated an
exclusion model with a somewhat less stringent excluded-volume constraint
than in ASEP.  This weaker exclusion gives rise to an effective cooperativity
and thereby to complex density profiles similar to what we find.  As we shall
argue, the existence of these complex profiles does not depend on detailed
microscopic rules, but is rather a consequence of the underlying cooperative
interactions between particles.  When sufficiently strong, these interactions
leads to an inflection point in the current-density curve; this feature is
the minimum requirement for the complex density-profile dynamics.

\section{Cooperative Exclusion Model}

In the CE model, a particle can hop to its vacant right neighbor at a rate
$r$ that depends on the occupancy of the previous site (Fig.~\ref{model}):
\begin{equation*}
  r= \cases{ 1 & previous\ site\ occupied,\cr
    \lambda& previous\ site\ vacant,}
\end{equation*}
with $0\leq \lambda \leq 1$.  When $\lambda=1$, the standard ASEP is
recovered, while $\lambda=0$ corresponds to \emph{facilitated} asymmetric
exclusion~\cite{BM09}, in which the left neighbor of a particle must be
occupied for the particle to hop to a vacancy on the right.  We pictorially
view this restriction as a particle requires a ``push'' from its left
neighbor to hop.  This facilitation causes an unexpected discontinuity in a
rarefaction wave in the ASEP~\cite{GKR10}.  More strikingly, we will show
that cooperativity leads to shock and rarefaction waves that can be
continuous, discontinuous, or a mixture of the two.

\begin{figure}[htb]
\begin{center}
\includegraphics[width=0.45\textwidth]{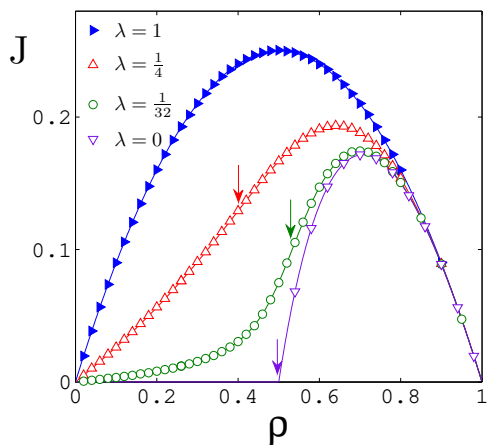}
\caption{Steady-state current as function of density in cooperative exclusion
  (CE).  Data are based $10^2$ realizations with $L=10^3$ up to $t=10^4$.
  The solid curves are given by Eq.~(\ref{current}).  Arrows indicate the
  locations of the inflection points.}
  \label{JvsRHO}
\end{center}
\end{figure}

These unusual features arise in CE when $0<\lambda< \frac{1}{2}$, where an
inflection point in $J(\rho)$ occurs at $\rho=\rho_I$ (Fig.~\ref{JvsRHO}).
For $\rho<\rho_I$, cooperativity dominates, and $J$ grows superlinearly in
$\rho$.  At higher densities, excluded volume interactions dominate, so that
$J$ grows sublinearly and ultimately decreases to zero.  Correspondingly, the
group velocity changes from an increasing to a decreasing function of density
$\rho$ as $\rho$ passes through $\rho_I$.

\begin{figure}[htb]
\begin{center}
  \includegraphics[width=0.3\textwidth]{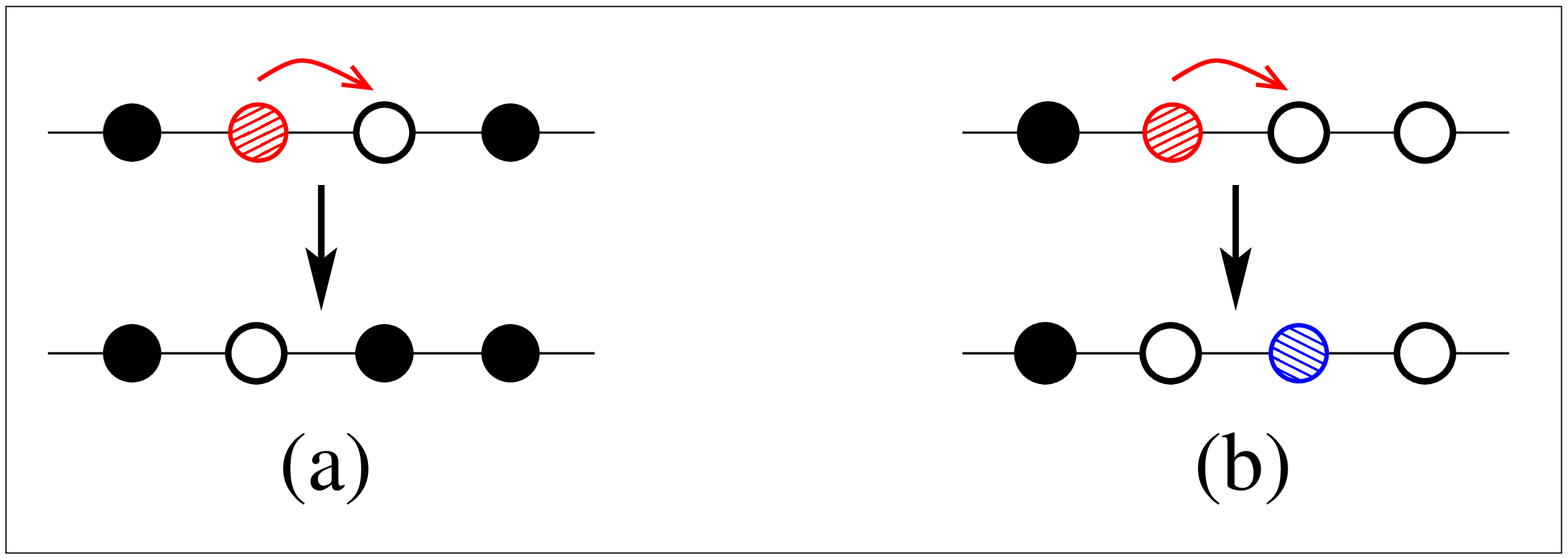}\qquad\qquad \includegraphics[width=0.3\textwidth]{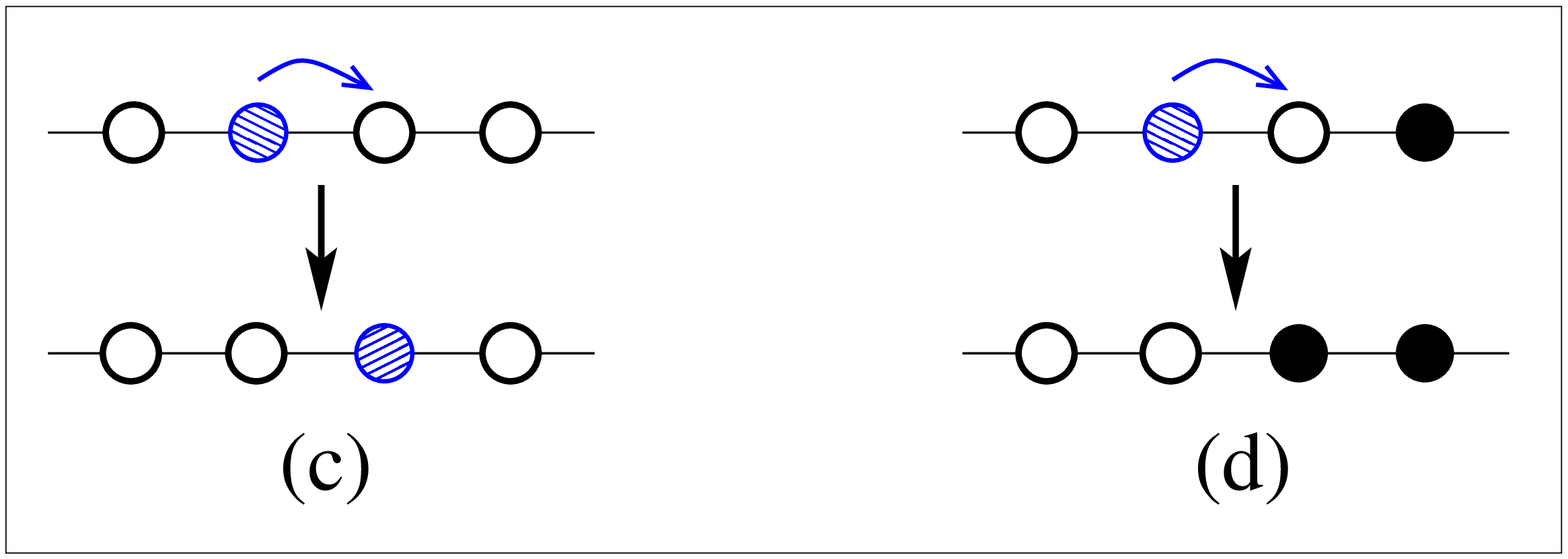}
\caption{[left] Hopping of a pushed (red) particle where the number of
  vacancy pairs is (a) preserved or (b) decreases.  [right] Hopping of an
  isolated (blue) particle where the number of vacancy pairs is (c) preserved
  or (b) increases. }
  \label{k}
\end{center}
\end{figure}

A configuration of $N$ particles on a ring of length $L$ is specified by the
occupation numbers $\{n_1,\dots,n_L\}$, subject to conservation $\sum_i
n_i=N$; here $n_i$ equals $1$ if $i$ is occupied and equals 0 otherwise.  A
crucial feature of CE is that the probability for any steady-state
configuration is a \emph{decreasing} function of the number $k$ of adjacent
vacancies: $k\equiv \sum_{i=1}^{L} (1-n_i)(1-n_{i+1})$, with $n_{L+1}=n_1$.
To understand how the configurational probabilities depend on $k$, we observe
that the hopping of a pushed particle (whose left neighbor is occupied)
either preserves or decreases the number of adjacent vacancies $k$
(left side of Fig.~\ref{k}).  Conversely, the hopping of an isolated particle
either preserves or increases $k$ (right side of Fig.~\ref{k}).  Since pushed
particle hopping events occur at a higher rate, configurations with fewer
adjacent vacancies are statistically more probable.

We now exploit the work of Antal and Sch\"utz~\cite{AS00} who investigated a
dual model in which next-nearest neighbor cooperative interactions pull a
particle ahead, in distinction to the pushing of particles from behind in CE.
By the mapping particles $\leftrightarrow$ holes, the CE and the
Antal-Sch\"utz models give the same probability distribution $P_k$ for a
configuration with $k$ adjacent vacancies~\cite{AS00}:
\begin{equation}
\label{prob}
P_k=\frac{\lambda^k}{Z(\lambda)}~,
\end{equation}
where $Z(\lambda)$ is a normalization constant.  Since $\lambda<1$,
configurations with fewer adjacent vacancies are more probable.
Following~\cite{AS00}, the steady-state current is
\begin{equation}
\label{current}
J=(1-\rho)\left[1+\frac{\sqrt{1-4(1-\lambda)\rho(1-\rho)}-1}{2(1-\lambda)\rho} \right]
\end{equation}
in the $L\to\infty$ limit.  The salient feature is that $J$ has an inflection
point at a density $\rho_I$ for $\lambda<\frac{1}{2}$ (Fig.~\ref{JvsRHO}).
We henceforth restrict our analysis to this domain and determine the unusual
consequences of this inflection point on the dynamics of initial density
steps.

\section{Density Profile Dynamics}
\label{dynamics}

\begin{figure}[ht]
\begin{center}
\includegraphics[width=0.35\textwidth]{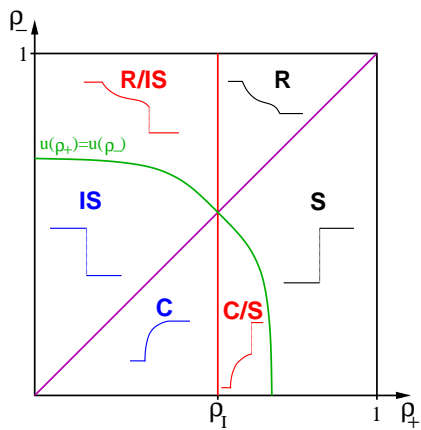}
\caption{Phase diagram of the CE model for an initial density step
  $(\rho_-,\rho_+$), with $\rho_I$ the inflection point in $J(\rho)$.  A
  typical density profile $\rho(z)$ is sketched for each of the six regions:
  (R/IS) rarefaction/inverted shock, (R) continuous rarefaction, (S) shock,
  (C/S) compression/shock, (C) continuous compression, (IS) inverted shock. }
  \label{phase_diagram}
\end{center}
\end{figure}

In a hydrodynamic description, the particle satisfies the continuity equation
$\rho_t + J_x= 0$.  By the chain rule, we rewrite the second term as
$J_\rho\,\rho_x$, from which the group velocity $u=J_\rho$.  Here the
subscripts $t,x,\rho$ denote partial differentiation.  The crucial feature is
the inflection point in $J(\rho)$, so that the group velocity can be either
increasing or decreasing in $\rho$.  We now employ the steady-state current
(\ref{current}) to determine the evolution of an initial density
heterogeneity on length and time scales large compared to microscopic scales
for the step initial condition
\begin{equation}
\label{initial_density}
\rho(x,t=0)= 
\cases {
  \rho_- & $x\leq0$\,, \cr
  \rho_+ & $x>0$\,.}
\end{equation}
As sketched in Fig.~\ref{phase_diagram}, the difference in the group
velocity to the right and left of the step determines whether a continuous,
discontinuous, or a composite density profile emerges.

It is worth noting that similar results for density profiles is obtained for
an asymmetric exclusion process with another form of cooperative
interactions~\cite{FGRS02,BGRS02}.  In these works, the same qualitative
phase diagram as in Fig.~\ref{phase_diagram} is obtained, despite the rather
different natures of the microscopic interactions in their model.  This
similarity in long-time behavior arises because our main results apply for
{\emph any} asymmetric exclusion process with sufficiently strong cooperative
interactions, as indicated by an inflection point in $J(\rho)$.

\begin{figure}[ht]
\begin{center}
\mbox{\subfigure{\includegraphics[width=0.4\textwidth]{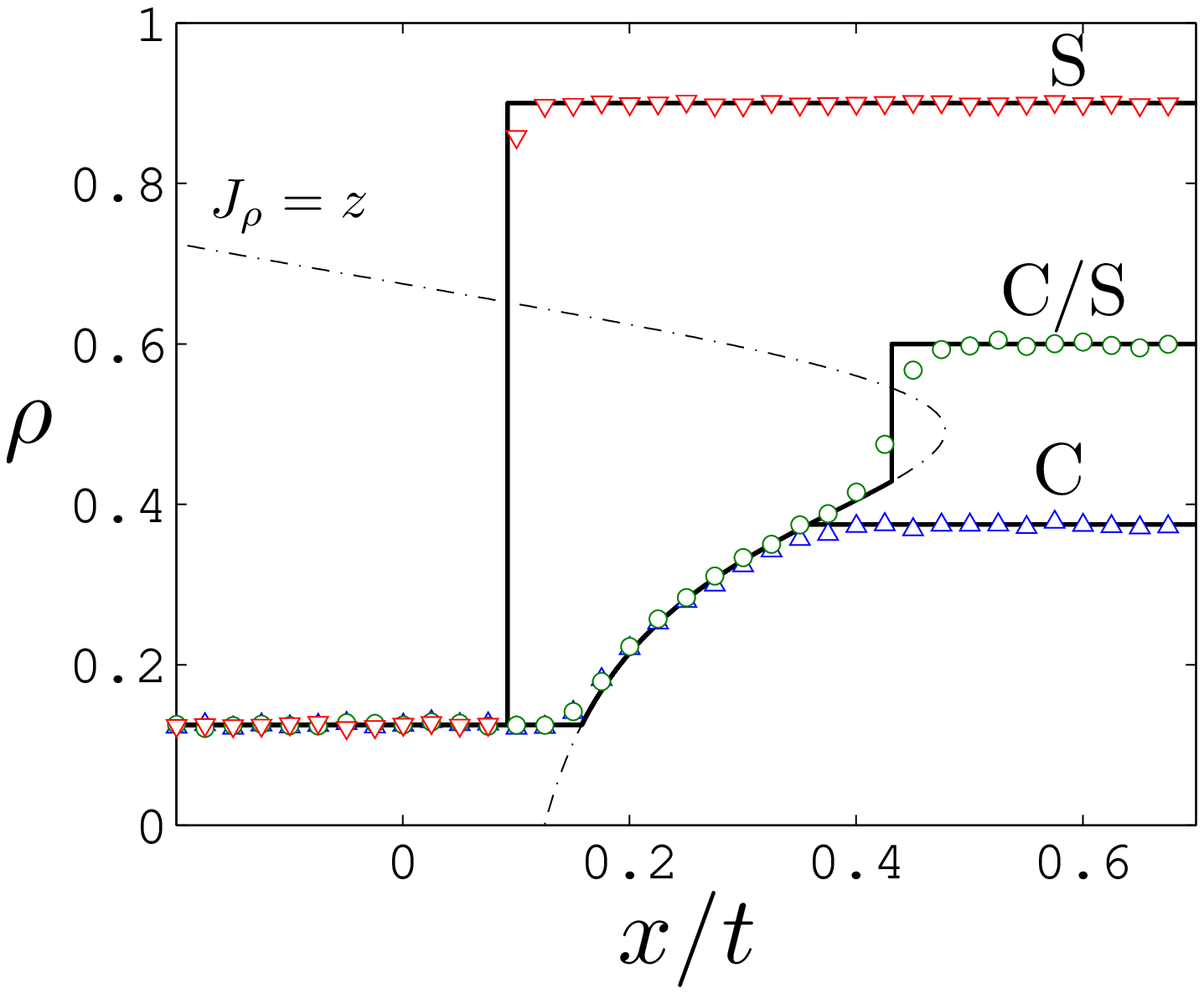}} \quad
\subfigure{\includegraphics[width=0.4\textwidth]{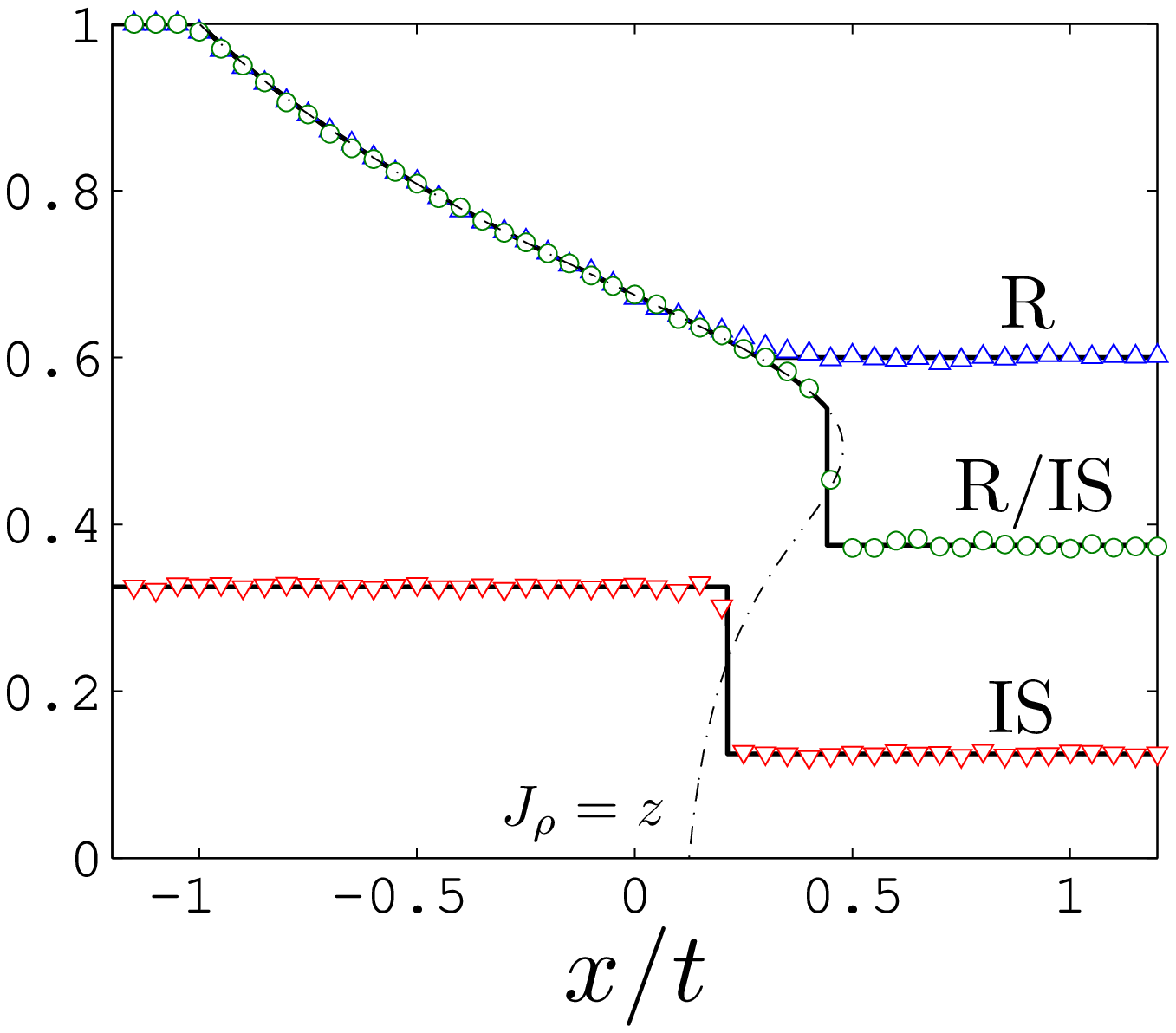}}}
\caption{(left) Evolution of an upstep for $\lambda=\frac{1}{8}$: (C)
  continuous compression wave for $\rho_-=\frac{1}{8}$, $\rho_+=\frac{3}{8}$;
  (C/S) composite compression/shock for $\rho_-=\frac{1}{8}$,
  $\rho_+=\frac{6}{10}$; (S) shock for $\rho_-=\frac{1}{8}$,
  $\rho_+=\frac{9}{10}$.  (right) Evolution of a downstep for
  $\lambda=\frac{1}{8}$: (R) continuous rarefaction for $\rho_-=1$,
  $\rho_+=\frac{6}{10}$; (R/IS) composite rarefaction/inverted shock for
  $\rho_-=1$, $\rho_+=\frac{3}{8}$; (IS) inverted shock for $\rho_-=0.325$,
  $\rho_+=\frac{1}{8}$.  The dashed line is the locus $J_\rho=z$ and the
  solid black curves are analytic predictions.  Simulations are based on
  $10^3$ realizations up to $t=4\times10^3$.}
  \label{steps}
\end{center}
\end{figure}

\emph{Shock/Inverted Shock:} A propagating shock wave arises whenever the
group velocity on the left exceeds that on the right, $u(\rho_-)>u(\rho_+)$.
Qualitatively, the faster moving particles catch up to slower particles on
the right and pile up in a shock wave, just as freely-moving cars suddenly
slow down upon approaching a traffic jam.  In the conventional ASEP, all
upsteps evolve into a \emph{shock} (S) wave.  For the CE, in contrast, only
upsteps where both initial densities are above the inflection
point, $\rho_I<\rho_-<\rho_+$, evolve into shocks (Fig.~\ref{steps}).  Here,
exclusion is sufficiently strong that the group velocity is a decreasing
function of density.  Strikingly, a propagating shock wave also emerges from
a downstep in CE when the initial densities are both below the
inflection point, $\rho_I>\rho_->\rho_+$.  In this regime,
$J_{\rho\rho}=u_\rho>0$; that is, cooperativity is sufficiently strong that
particles in the high-density region on the left have a greater group
velocity and therefore pile up at the interface.  We term this singularity an
\emph{inverted shock} (IS) (Fig.~\ref{steps}).

For both shocks and inverted shocks, the density is given by the traveling
wave profile $\rho=\rho(x-ct)$.  We obtain the shock speed $c$ by equating
the net flux into a large region that includes the shock,
$J(\rho_+)-J(\rho_-)$, with the change in the number of particles,
$c(\rho_+-\rho_-)$, in this region~\cite{W74} to obtain the standard
expression $c = [{J(\rho_+)-J(\rho_-)}][{\rho_+-\rho_-}]$; this holds both
for conventional and inverted shocks.

\smallskip \emph{Continuous Rarefaction/Compression:} A density step
gradually smooths out when the when the group velocity to the left is less
than that on the right, $u(\rho_-)<u(\rho_+)$.  Here the faster particles on
the right leave open space for the slower particles, similar to a cluster of
stopped cars that slowly spreads out after a stoplight turns green.  In ASEP,
a downstep always evolves to a \emph{continuous rarefaction} (R) wave.  This
continuous rarefaction also occurs in CE when both initial densities are
above the inflection point, $\rho_->\rho_+>\rho_I$.  At these high densities,
exclusion dominates, as in the ASEP, which causes the group velocity to
decrease with density.

In striking contrast to the ASEP, an upstep can continuously smooth out in CE
when the initial densities are below the inflection point,
$\rho_-<\rho_+<\rho_I$.  In this regime, cooperativity is sufficiently strong
that particles in the high density region on the right move faster than those
on the left.  Thus instead of a shock wave, a \emph{continuous compression}
(C) wave develops (Fig.~\ref{steps}).  We determine the density profile by
assuming that it a function of the scaled variable $z=x/t$.  Substituting
$\rho(x,t)=\rho(z)$ into the continuity equation gives
$-z\rho_z+J_\rho\,\rho_z=0$.  Thus the profile consists either of
constant-density segments ($\rho_z=0$) or else $z=J_\rho$.  Matching these
solutions gives~\cite{KRB10,GKR10}
\begin{equation}
\label{rarefaction}
\rho(z)= \cases{ \rho_- & $z<z_-$ \,,\cr
I(z) & $z_-\leq z \leq z_+$\,, \cr
\rho_+ & $z>z_+$ \,,}
\end{equation}
where $I(z)$ is the inverse function of $z=J_\rho$.  For a continuous
profile, the cutoffs $z_-$ and $z_+$ are determined by matching the interior
solution $I(z)$ with the asymptotic solutions: $I(z_\pm)=\rho_\pm$ or
equivalently, $z_\pm=J_\rho(\rho_\pm)$.

\smallskip \emph{Composite Rarefaction/Compression and Shock:} In CE, a
continuous rarefaction or compression wave can coexist with a shock wave.
This phenomenon occurs when the group velocity on the left is initially less
than that on the right but also with the constraint that the initial
densities lie on either side of the inflection point.  Consequently one side
of the step is in the exclusion-dominated regime and the other is in the
cooperativity-dominated regime, or vice-versa.  In particular, a
\emph{composite rarefaction/inverted shock} (R/IS) wave emerges from a
downstep when $\rho_->\rho_I>\rho_+$, so that $u(\rho_-)<u(\rho_+)$.  As in
the case of the continuous rarefaction wave, the downstep begins to smooth
out from the rear.  Consequently, cooperative interactions become more
important as the density at the leading edge of this rarefaction decreases.
Eventually this leading density reaches the point where the particle speed
matches that at the bottom of the downstep and the rarefaction front
terminates in an inverted shock.

Correspondingly, an upstep can evolve to a compression wave with a leading
shock when the densities satisfy $\rho_-<\rho_I<\rho_+$ and
$u(\rho_-)<u(\rho_+)$.  In this case, the leading particles initially race
ahead, leaving behind a profile where the density increases with $x$.
However, this increase cannot be continuous because eventually a point is
reached where the speed at the front of this continuous wave matches that of
the top of the upstep.  After this point, a pile-up occurs and a shock wave
forms.  We call this profile a \emph{composite compression/shock} (C/S) wave
(Fig.~\ref{steps}).

The functional forms of the composite rarefaction/inverted shock and
composite compression/shock profiles are still given by
Eq.~(\ref{rarefaction}), but the criteria to determine the cutoffs $z_\pm$
are now slightly more involved than for continuous profiles.  The location of
the left cutoff, $z_-$, is again determined by continuity, namely,
$I(z_-)=\rho_-$ or, alternatively, $z_-=J_\rho(\rho_-)$.  To determine the
right cutoff $z_+$, note that in a small spatial region that includes the
leading-edge discontinuity, the density profile is just that of a shock or
inverted shock wave.  Thus the equation for the shock speed is
\begin{equation}
\label{interface}
z_+=\frac{J(q_+)-J(\rho_+)}{q_+-\rho_+}~,
\end{equation}
where $q_+\equiv I(z_+)$ is the density just to the left of the
discontinuity.  (Note also that $z_+ =J_\rho(q_+)$ by definition.)  To
justify (\ref{interface}), we use the conservation equation that the particle
number in $[z_-,z_+]$ equals the initial number plus the net flux into this
region:
\begin{equation}
\label{particle_cons}
\int_{z_-}^{z_+}\!\! I(z)dz = -\rho_-z_-+\rho_+z_+-J(\rho_+)+J(\rho_-)\,.
\end{equation}
We recast this expression into (\ref{interface}), by making the variable
change $z=J_\rho(\rho)$ and using $I(J_\rho(\rho))=\rho$ to write the
integral as $\int_{\rho_-}^{q_+}\rho\,J_{\rho\rho}\,d\rho$, which can be
performed by parts.  The resulting expression readily simplifies to
(\ref{interface}). 

In summary, a diversity of wave singularities arise in asymmetric exclusion
with sufficiently strong cooperativity.  The minimum requirement for these
phenomena is an inflection point in the current-density relation $J(\rho)$.
This inflection point leads to a group velocity that is an \emph{increasing}
function of density for $\rho<\rho_I$, a dependence opposite to that in the
conventional ASEP.  The resulting non-monotonic density dependence of the
velocity causes an initial density upstep or downstep to evolve to:
shock/inverted shocks, continuous rarefaction/compression waves, or a
composite profile with both continuous and discontinuous elements.

\ack We thank Martin Schmaltz for asking an oral exam question that helped
spark this work and Paul Krapivsky for helpful discussions.  We also thank
the referee for informing us about Refs.~\cite{FGRS02,BGRS02}.  Finally, we
gratefully acknowledge financial support from NSF grant DMR-0906504.

\end{document}